\documentclass[epj]{svjour}
\usepackage{graphicx}
\usepackage{amsmath,amssymb}

\begin{document}
\title{Granular size segregation in underwater sand ripples}
\author{G. Rousseaux\inst{1}, H. Caps\inst{2} and J.-E. Wesfreid\inst{1}}
\institute{Physique et M\'ecanique des Milieux H\'et\'erog\`enes (PMMH),
UMR CNRS-ESPCI 7636- 10 Rue Vauquelin - 75231 Paris Cedex 05
(France) \and Group for Research and Applications in Statistical Physics (GRASP), Universit\'e de Li\`ege - Institut de
Physique B5 - B-4000 Li\`ege (Belgique)}

\abstract{We report an experimental study of a binary sand bed under an oscillating water flow. The formation and evolution of ripples is observed. The appearance of a granular segregation is shown to strongly depend on the sand bed preparation. The initial wavelength of the mixture is measured. In the final steady state, a segregation in volume is observed instead of a segregation at the surface as reported before. The correlation between this phenomenon and the fluid flow is emphasised. Finally, different ``exotic'' patterns and their geophysical implications are presented.}

\maketitle

\section{Introduction}

On a sand bed eroded by a fluid, such as air or water, ripple formation is generally observed \cite{ball}. Coastal areas as well as desert landscapes are covered by these structures. Despite their familiar aspect, the physical mechanisms involved are related to complex granular transport processes, as described by Bagnold \cite{Bagnold41,Bagnold46}. The underwater sand ripples observed at the beach are due to the elliptical motion of the water particles underneath the surface waves \cite{Guyon}. The shear stress applied to the sand grains by the water imposes an oscillating motion of the grains. The pattern formation is a spontaneous consequence of the grain motion. 

Natural sand beds are generally composed of different granular species. Broad granulometric distributions are indeed observed \cite{hunter}. As a consequence, phase segregation and stratigraphy may occur \cite{hunter,Ottino,Doucette,Seminara,FB,Makse,caps}.

In this paper we propose an experimental study of underwater sand ripples formed with a binary granular mixture. In Section II, we present the experimental setup. The impact of the initial preparation of the sand bed on the patterns is emphasised in Section III. Then, in Section IV, our experimental results are presented. A rapid review of different exotic segregation patterns is proposed in Section V. Eventually, a summary of our findings is given in Section VI.

\section{Experimental setups and procedures}

In this section, we first present the experimental setup. Then, we focus on hydrodynamics in order to justify some experimental assumptions.

\subsection{Experimental geometry}

\begin{figure}[ht]
\begin{center}
\includegraphics[width=7cm]{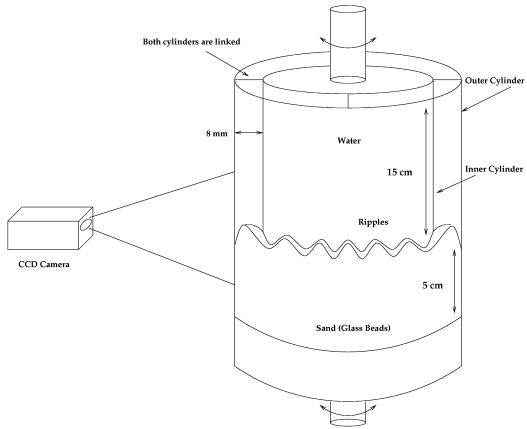}
\caption{Experimental setup for the ripple formation and analysis
(Setup 1).} \label{dispo}
\end{center}
\end{figure}

In this study, we used two experimental setups. Both setups are composed of two concentric cylinders joined by two circular plates on the top and on the bottom (for further details, see \cite{Stegner}). The so-created annular channels are fully filled with a layer of sand ($\approx 5$ cm) and water [see Figure~\ref{dispo}]. The mean radii of the setups are $7.31$ cm and $13.55$ cm for Setup 1 and Setup 2 respectively. The width of channels are $0.8$ cm and $1.9$ cm respectively. With the help of a motor, the sand bed is oscillated at fixed amplitude $A$ and frequency $f$. Herebelow, $A$ ranges from $1.5$ cm to $5$ cm and $f$ ranges from $0.5$ Hz to $4$ Hz. After several oscillations, ripples appear all around the perimeter. 

The main advantage of such an annular geometry, compared to linear channels \cite{Foti}, is a strict mass conservation and periodic boundary conditions. 

In Setup 1, a fixed CCD camera is horizontally placed at the same height as the sand/water interface and records images of the interface. Each acquisition is performed by driving the setup at low velocity and recording profiles continuously. For setup 2, a CCD camera is placed on the top of the setup and records images of the whole interface reflected on a conical mirror. In addition to this, a camera can be fixed on the oscillating plate in order to record side views of the interface evolution in its rotating frame.

We used spherical glass beads and glass powders grains of different sizes and colours with a density $\rho_s=2.5$ kg/l. The different sizes are obtained by sifting so that the silver blue spherical beads have a mean size $315<d<355\,\mu$m, while the silver yellow ones are characterised by $250<d<280\,\mu$m. The grains of one powder are black and have a size $250<d<350\,\mu$m, while the small yellow ones have $150<d<250\, \mu$m. We used two binary mixture : (i) 50\% in volume of big blue beads and 50\% in volume of small yellow beads for mixture $M1$ and (ii) 50\% in volume of big black grains and 50\% in volume of small yellow grains for mixture $M2$.

\subsection{Hydrodynamic considerations}

One should note that in our experiments, the sand bed is oscillating under steady water, contrary to the beach case. In order to ensure that this procedure is equivalent to the natural case, some care has to be taken. Indeed, as the channel is moving relative to the fluid, velocity is diffused in the fluid by viscous effects. These effects occur within a typical length $\delta$ - called {\it Stokes layer} - which can be written as \cite{Guyon}
\begin{equation}
\delta = \sqrt{\frac{\nu}{\pi f}},
\end{equation}
where $\nu$ is the kinematic viscosity of the fluid.

At a distance from the wall less than the Stokes layer, the flow is close to a uniform shearing motion \cite{Bagnold46}. The grains there are submitted to a uniform and oscillating shear stress similar to the back and forth motion induced by sea waves on the beach. Thus, the widths of our experimental channels (typically $1$ cm) have been chosen in such a way that the Stokes layer caused by the vertical walls of the channel (typically $560\, \mu$m at 1 Hz)  can be neglected. At the sand-water interface, the same consideration applies since the grain sizes does not exceed $355\, \mu m$ in our experiments.



It can also be useful to evaluate the laminar {\it Shields parameter} $\Theta_L$ which is the ratio between the shear forces caused by a laminar flow on one grain and the gravity force acting on it \cite{Bagnold46}. With the previous notation, $\Theta_L$ reads

\begin{equation}\label{shield}
\Theta_L = \frac{\tau_L d^{2}}{(\rho _{s}-\rho _{f}) g d^{3}}
\simeq \frac{\rho _{f}\nu V }{\Delta \rho g d \delta} \sim
\frac{\rho _{f}\nu^{1/2} A f^{3/2} }{\Delta \rho g d},
\end{equation}
where $\tau_L$ is the shear stress and $V$ is the fluid velocity. The condition for grain motion is thus $\Theta_L>\Theta_c$, with $\Theta_c\approx 0.05$ following Nielsen \cite{Nielsen92}. But the estimate of this value is actually subject of study. One should note that even the ripples modify the flow and may create vortices during small parts of the oscillations. The flow can be considered as laminar during the remainder part until the value of the Reynolds number is less than $100$ \cite{jensen}.

\section{Interface preparation}

In order to study phase segregation, special care needs to be paid in the generation of a uniformly mixed initial configuration. As the sand bed flattening is performed by a stage of oscillations at large $A$ and $f$ values, phase segregation may occur during this stage. A systematic and reproducible procedure is thus needed to flatten the sand bed prior to each experiment. However, it should be noticed that the quantitative study of the response of a granular bed to strong oscillations is out of the scope of this paper. What we present here are mainly qualitative results coming from several observations in many experimental conditions.

\begin{figure*}[htb]
\begin{center}
\includegraphics[width=15cm,height=3cm]{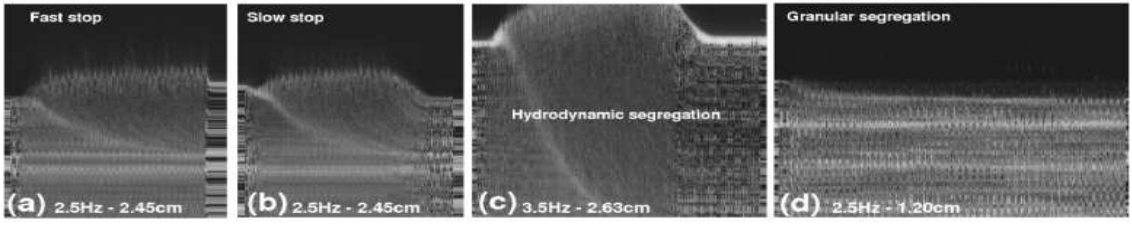}
\caption{Spatio-temporal side views of the sand bed before, during and after the preparation procedure. Time goes from left to right. (a) A fluidization step followed by a brutal stop leads to a loose packed sand bed ; (b) a fluidization step followed by a slow stop leads to an homogeneous and compacted mixture ; (c) hydrodynamic phase segregation (small white grains on the top) appears if the grains are put into suspension ; (d) granular phase segregation (large black grains on the top) results from the Brazil-nut effect caused by a small oscillation of the setup. All pictures are 1.1 cm height and 50 s width and are performed with mixture $M2$ in setup 1.}\label{segr_prep}
\end{center}
\end{figure*}

First, if the sand bed is strongly oscillated such that the grains are put into suspension and then left at rest, the bigger grains are observed to fall down much more rapidly than the smaller grains, according to the sedimentation law \cite{Guyon} :
\begin{equation}
v_{s}=\frac{\Delta \rho g d^{2}}{18 \rho_f\nu},
\end{equation}
where $v_{s}$ is the sedimentation velocity of isolated grains. This difference in sedimentation velocities results in a larger amount of small grains on the top of the sand layer. We call this process an {\it hydrodynamic phase segregation}. Figure~\ref{segr_prep} (c) presents a spatio-temporal diagram of a sand bed prepared in such a way ($A=2.63$ cm and $f=3.5$ Hz). The first growth of the sand/water interface is due to the dilatancy caused by the oscillations, while the decrease comes from the sedimentation. It can be noticed that this kind of hydrodynamic segregation is not specific to laboratory experiments but may also occur in nature during storms \cite{Doucette}.

Secondly, if the oscillations are such that the sand bed is fluidized but without suspension, we observe that the depth of movable material increases with both $A$ and $f$. In those situations, the larger grains are seen to move up in the sand bed, as in the Brazil nut effect \cite{Leonardo}. We qualify this phenomenon of a {\it granular phase segregation}. This results in a large amount of larger grains on the top of the sand bed, as illustrated in Figure~\ref{segr_prep} (d). In this figure, the sand bed dilatancy is small, due to the lower values of $A$ and $f$.

The compaction of the bed must also be considered. We have noticed that a loose bed is more easily penetrated by the fluid. In this case, during the first oscillations, the fluid is observed to carry layers of grains instead of individual particles. However, one of our aims is to observe the very first steps of the ripple formation, when the ripples are composed of a few grains. Therefore, we have decided to start from a compacted bed and avoid transient effects. Experimentally, the way the sand bed is compacted depends on the way the oscillations are stopped. Abruptly stopping the oscillations causes a smaller compaction of the pile than a slow decrease of $A$ or $f$. More precisely, a brutal stop leads to a compaction smaller than those of a disordered packing ($\rho\approx 0.6$ \cite{Ona}). This behaviour is clearly emphasised in Figure~\ref{segr_prep} (a) where one can see that the sand/water interface is higher after the stop than before the oscillations. 

The procedure we used is largely empirical but leads to highly reproducible results. We first oscillate the setup at a high frequency $f$ and amplitude $A$ values. This destroys the former patterns. Then, we decrease the amplitude $A$ of oscillation keeping the frequency $f$ constant and we let the setup oscillate for a while with these $A$ and $f$ parameters. This increases the compaction of the pile. The setup oscillates until no more decrease of the sand/water interface is observed. Thus although we don't know quantitatively the value of the density of the pile, we are sure that we are in the ``most'' compacted state. The time required for reaching this stable state has been evaluated for different grain sizes : $2$ s for $d=300\, \mu$m, $15$ s for $d=110\, \mu$m and $60$ s for $d=65\, \mu$m. This time is thus growing when the grain size decreases.

The measurement of the homogeneity of the mixture is a very hard task due to the 3D property of our packing. However, we have observed that strong oscillations of the setup lead to a slight hydrodynamic segregation which is counterbalanced by granular segregation occurring during the compaction step. Nevertheless, our results seem to be reproducible as far as the preparation procedure is used.

Eventually, we have noticed that during the fluidization of the sand/water interface transients Kelvin-Helmholtz waves appear at the separation between the fluid and the fluidized bed, as previously observed \cite{Evesque}.


\section{Results}
In this section, we present the experimental results concerning the initial stage of ripple formation in binary mixtures. Then, we study the phase segregation observed in the stable state.

\subsection{Wavelength selection}

The destabilisation of the sand/water interface is characterised by the appearance of small bumps at different places on the interface. These are called rolling-grain ripples \cite{Bagnold46}. We define the {\it initial wavelength} $\lambda_0$ of the ripples as the width of the bump base. 

\begin{figure}[hbt]
\begin{center}
\includegraphics[width=8cm]{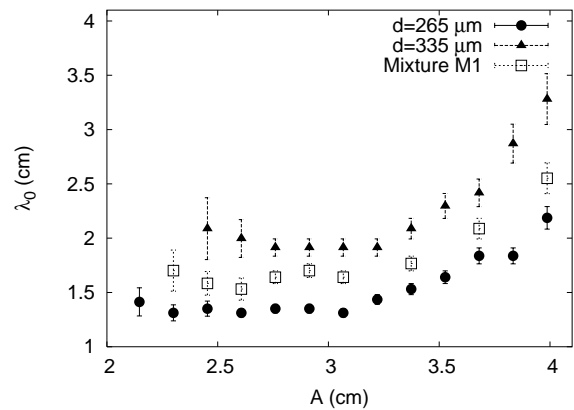}
\caption{Initial wavelength $\lambda_0$ as a function of the amplitude $A$ of oscillations for granular species with sizes $315<d<355\,\mu m$ and $250<d<280\,\mu m$ as well as for the mixture of the two species. Frequency $f=1$ Hz.}\label{init}
\end{center}
\end{figure}

The coalescence of rolling-grain ripples is fast and occurs as soon as two of them are close to each other. The measurement of $\lambda_0$ is thus complicated. Fortunately, the ripple coarsening depends on the acceleration $Af^2$. For small values of the acceleration, the coarsening is slow and $\lambda_0$ can be measured accurately. Figure~\ref{init} presents the initial wavelength $\lambda_0$ of the patterns as a function of the amplitude $A$ of oscillations for a frequency $f=1$ Hz in the case of mixture $M1$. The curves corresponding to the separated species composing $M1$ are also illustrated \cite{Rousseaux}.  

In all cases, $\lambda_0$ exhibits a constant regime followed by a slight growth for $A>3.1$ cm. This growth may be affected by ripple coalescence. Indeed, at fixed $f$ the ripple coalescence appears as soon as $A$ is large. This also applies for $A$ constant and $f$ varying. Therefore, the rolling-grain state may be by-passed towards the regime characterised by larger structures, the vortex-ripples. In such a case, the measurement of the initial wavelength is of course difficult. We have thus considered that the initial wavelength $\lambda_0$ is independent of the amplitude $A$ of oscillations at fixed frequency $f$ close to the threshold for grain motion. We have also observed that $\lambda_0$ does not depend on the frequency $f$ at fixed amplitude $A$ \cite{Rousseaux}.

An important observation is that $\lambda_0$ increases with the grain size $d$ and, moreover, that the mixture behaves in an intermediate way between the species composing it. The curve corresponding to the mixture is similar to that of a monodisperse species of size $d=(265+335)/2\, \mu$m. If we consider that the amount of grains composing the primary structures is nearly constant, then the growth of $\lambda_0$ with $d$ may be seen as trivial. On should also notice that the data corresponding to the lowest $A$ value of each curve is the minimum amplitude required for ripple formation at $f=1$ Hz. The value of this threshold increases with the grain size, which is consistent with the dependence of the Shields number on the ratio $A/d$ [cfr Eq.(\ref{shield})].

All these mainly qualitative results clearly show that the morphology of the primary ripples is governed by granular phenomena. 

\subsection{Final steady state}

The coalescence of rolling-grain ripples results in larger structures called {\it vortex-ripples} \cite{Bagnold46}. After several hours, the sand/water interface reaches a steady state characterised by a constant ripple wavelength and a constant ripple amplitude $h$. We have performed several experiments with various amplitude $A$ and frequency $f$ of oscillations in different granular mixtures. The mean wavelength $\lambda_\infty$ and the mean ripple amplitude $h$ of the vortex-ripples are measured with the help of Fourier series decompositions. 

\begin{figure}[!h]
\begin{center}
\includegraphics[width=8cm]{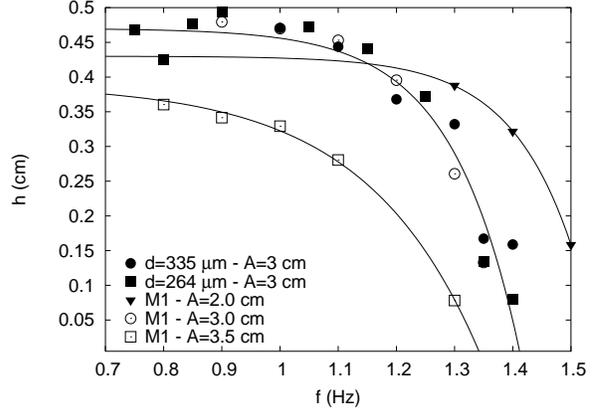}
\caption{Mean ripple amplitude $h$ as a function of the oscillation frequency $f$ for different grains sizes $d=265 \mu m$, $d=335 \mu m$ and for mixture M1. Different amplitude values are illustrated. Solid lines are guides for the eyes.}\label{hfcomp}
\end{center}
\end{figure}

\begin{figure*}[htbp]
\begin{center}
\includegraphics[width=15cm,height=7cm]{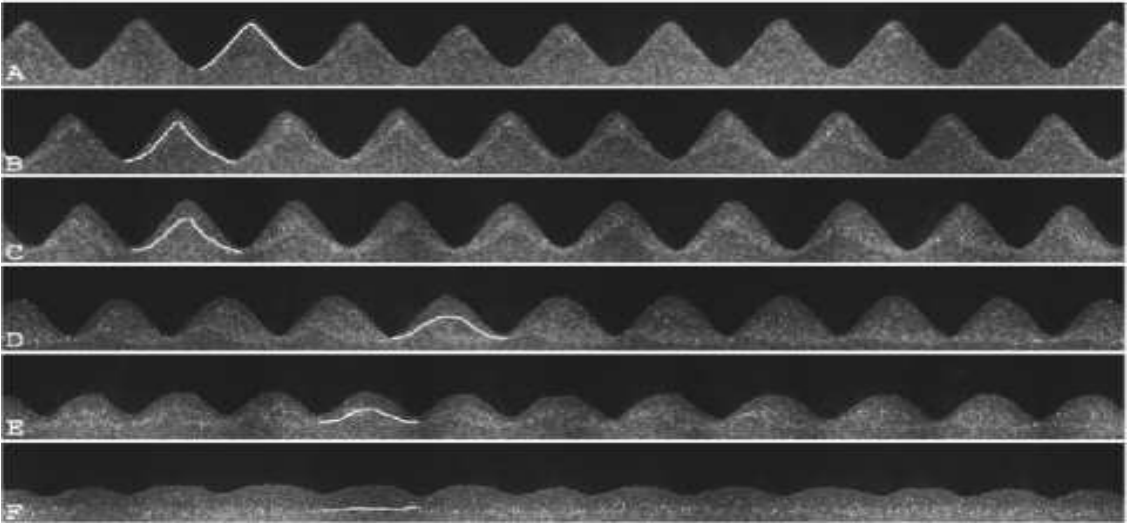}
\caption{Side views of the final steady state (Mixture $M1$ in Setup 1). Different frequencies are illustrated ($A=3$cm): (a) $f=0.8\, Hz$, (b) $f=0.85\, Hz$, (c) $f=0.9\, Hz$, (d) $f=1\, Hz$, (e) $f=1.1\, Hz$, (f) $f=1.2\, Hz$. Solid white lines emphasise the limit of the segregated zone. The vertical scale is stretched 2.5 times with respect to the horizontal.}\label{binImage}
\end{center}
\end{figure*}

We have observed that the final wavelength $\lambda_\infty$ depends neither on the frequency $f$ of oscillations nor on the grain size $d$, but depends on the amplitude $A$ of oscillations. This is also the case in monodisperse sand beds \cite{Stegner}. Figure~\ref{hfcomp} shows that the mean ripple amplitude $h$ decreases when the frequency $f$ increases. This is also emphasised by Figure~\ref{binImage}, where images of final configurations are presented for different frequencies of oscillation $f$. This decrease of $h$ is due to the strength of the fluid flow at large $f$ values that flattens the ripple crests. In all cases, this flattening seems not to be of inertial forces nature since all the grains should be in relative motion to the walls in such a case. Nevertheless, we would like to stress that these observations show us that the vortex-ripple morphology is mainly controlled by hydrodynamic effects, whatever the poly-dispersity of the bed.   

\begin{figure}[ht]
\begin{center}
\includegraphics[width=9cm,height=3cm]{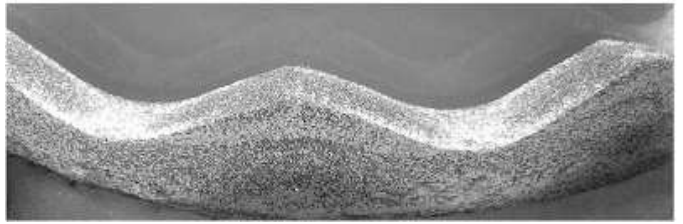}
\caption{Image of three ripples in their final steady state (Mixture $M1$ in Setup 2) with experimental parameters $f=0.5\, Hz$ and $A=6\, cm$. The larger grains (dark) are mainly found on the ripple crests while the smaller ones (white) are mainly located in the bottom.} \label{seg}
\end{center}
\end{figure}

Starting from an homogeneous mixture, the system always evolves through a segregated state. Looking from the top of the interface, one sees that a larger amount of large grains are found near the ripple crests while the smaller grains are mainly found in the bottom parts of the ripples [see Figure~\ref{seg}]. This observation was also reported in \cite{FB}. However, our experimental setup allows us to look inside the sand bed. Hence, we have observed that the {\it apparent}  segregation at the surface is actually the signature of a {\it segregation in volume}. Indeed, we have seen the formation of a thin layer of small (white) grains under the sand/water interface as presented in Figure~\ref{binImage}. This layer intercepts the sand/water interface at the bottom of the ripples, resulting in the misleading conclusion of a segregation at the surface. 

The segregation appearance is characterised by convective motions of both species from the crest to the troughs of the ripples, inside the ripples. As times goes on, the larger grains are repelled on both slopes, leaving the smaller ones in the bulk. One should note that this phase segregation in volume occurs on time-scales larger than the time needed to reach the final morphological state. The final steady state is thus declared when both $A$ and $\lambda_\infty$ are constant and when the layer of small grains does not evolve anymore. We also take care that no drift of the ripples is observed. Typically, those quantities must remain for times larger than one hour.  


As the frequency $f$ of oscillations increases [Figure~\ref{binImage} from (a) to (f)], the distance $d_s$ at the crests between the layer of small grains and the sand/water interface increases. In the bottom parts of the ripples, the layer remains nearly unchanged until the frequency $f$ becomes large enough to move all the layer deeper in the sand bed [Figure~\ref{binImage} (f)]. 

The layer of small grains can be seen as the bottom of a segregated zone which is larger near the crests and vanishes near the bottom parts of the ripples. Since the depth of this segregated zone strongly depends on the oscillation parameters, the fluid flow may be considered as the source of this layer appearance. We have thus numerically modelled our experimental setup in order to estimate the shear stress caused by the fluid flow on the sand bed. 

With the help of the finite-element code CASTEM from the CEA \cite{castem}, we have computed the Navier-Stokes equations \cite{Guyon} for a 2D water flow over a rippled wall. The wall shape is directly extracted from experiments. In order to take into account the setup oscillations in these stationary flow simulations, the shear stress $\tau$ during one oscillation is obtained by averaging the shear stresses corresponding to different velocities ranging from $0$ to $Af$. Since our computations do not account for time-dependence and flow separations, the predicted values for the shear stress may be lower than those expected with turbulent modelling \cite{anderson}. However, the deduced tendencies should be the same in both cases and stationary flows are more easily computed.
 


We have computed the shear stress $\tau$ over steady interfaces in different experimental conditions. We have observed that $\tau$ is larger for large ripples height $h$, i.e. for small frequency $f$ of oscillation. This was also mentioned in previous numerical studies \cite{kouame}. This means that the constraint imposed by the ripples on the fluid predominates over the shear stress due to fluid velocity.

\begin{figure}[ht]
\begin{center}
\includegraphics[width=7cm]{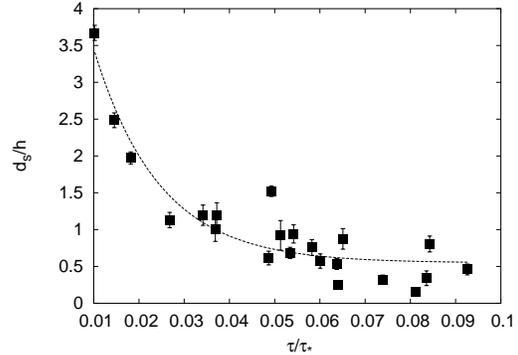}
\caption{Segregated zone depth $d_s$ at the ripple crests divided by the ripple height $h$ as a function of the shear stress $\tau$ at the sand/water interface normalised by the maximum plane shear stress $\tau_\star$.}\label{d_tau}
\end{center}
\end{figure}

Figure~\ref{d_tau} presents the distance $d_s$ at the ripple crests between the sand/water interface and the layer of small grains as a function of the shear stress $\tau$. Both quantities are respectively scaled by the ripple height $h$ and the theoretical maximum shear stress $\tau_\star=(\mu V)/\delta$ corresponding to a flat plane.

One can see that the ratio $d_s/h$ decreases as a function of $\tau/\tau_\star$ and then seems to saturate at a minimum value. We have assumed the following law
\begin{equation}\label{d_s}
\frac{d_s}{h}=a+b\,\exp\left(-\frac{\tau}{\tau_\star\, c}\right) ,
\end{equation}
where $a$, $b$ and $c$ are free fitting parameters with values $a=0.55\pm 0.11$, $b=5.79\pm 1.08$ and $c=0.011\pm 0.003$ respectively. It should be noticed that these parameters values may depends on the way $\tau$ is estimated but that the law should remain in all cases. 

The scenario for the segregation appearance may thus be the following. Due to its permeability, the sand-bed is penetrated by the velocity profile of the fluid. This leads to a ``Reynolds dilatancy'' that creates free spaces in the granular assembly. Differences of compaction between the crest and the trough of the ripple drive convective motions of both species towards the bottom. During this process, the smaller grains percolate down in the packing by means of those free spaces. The thin layer of small grains would thus correspond to the limit of the layer where convective motions (i.e. percolation) may occur. As the frequency $f$ increases, percolation is allowed on larger depths and the thin layer is expelled from the sand/water interface. Eventually, for very large $f$ values, the ripples are very flat and all the layer is well below the interface, giving large values for the ratio $d_s/h$. On the contrary, for small $f$ values, percolation occurs on small depths, giving small $d_s/h$ ratio values.

\begin{figure}[ht]
\begin{center}
\includegraphics[width=7cm]{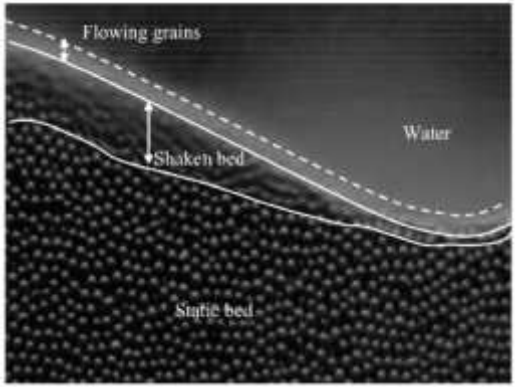}
\caption{Average of different images of a ripple during the fluid oscillations (Setup 2 and $d=500\, \mu m$ for visualisation purposes). The grains in motion are ``blurred'' while the static part of the pile is in focus.}\label{prof}
\end{center}
\end{figure}

As a proof of a segregation due to a ``shaking'' of the ripples by the fluid flow, we have recorded close-up movies of vortex ripples during oscillations. Averaging many images allows us to emphasise the motion of the sand grains [see Figure~\ref{prof}] :  a layer of grains in motion is observed over a layer of static grains. The depth of the moving grain layer is maximum near the ripple crests and decreases near the bottom parts of the ripples. As we increase the frequency $f$ of oscillations, the depth of the moving layer increases. This result clearly shows that the fluid stress penetrates the sand bed and allows grain motion. Figure~\ref{prof} refers to ripples created in a monodisperse ($d=500\, \mu m$) sand bed for clarity. If the sand bed is polydisperse, percolation is allowed and segregation in volume appears.  

One should eventually note that the decrease of the segregated zone depth from the ripples crest to the bottom is in agreement with the numerical calculations. Indeed, in all cases, the shear stress $\tau$ over the ripple is observed to be maximum near the ripple crest and then decreases near the bottom. Circulation zones may also be observed downstream from the ripples.

\section{Exotic patterns}

In this section, we present some special segregation features that often occur. We relate these observations to their geological implications. 

\subsection{Radial segregation}

\begin{figure}[ht]
\begin{center}
\includegraphics[width=8cm]{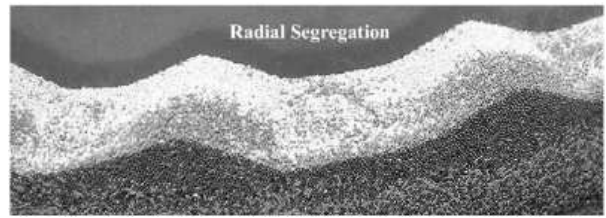}
\caption{Radial segregation is characterised by large grains preferentially situated close to the outer wall while the small ones go near the inner wall. Mixture $M1$ in Setup 2. frequency $f=1$ Hz and amplitude $A$2 cm.} \label{radial}
\end{center}
\end{figure}

Despite the nearly 2D geometry of our setup, we sometimes observe radial segregation. A larger amount of large grains are found near the outer cylinder while the smaller ones are preferentially found near the inner cylinder [see Figure~\ref{radial}]. We have not been able to determine exactly the experimental conditions leading to such a segregation. We first though that centrifugal forces were the key to this process : the larger grains would simply be moved further than the smaller ones. However, radial segregation preferentially occurs for small acceleration values (e.g. $A=2$ cm, $f=1$ Hz). For larger acceleration values (e.g. $A=3$ cm, $f=1$ Hz) radial segregation was absent. Moreover, this segregation takes a very long time to establish (typically $10$ hours) even though the final steady state was reached within one hour. Thus, we suspect this segregation to be an artifact due to the circular geometry of our setup. However, a deeper and systematic study is needed in order to explain quantitatively this result.

\subsection{Cat's eyes}

Before reaching the final steady state, the vortex ripple wavelength evolves through ripple coalescence and ripple annihilation processes \cite{Bagnold46}. Both processes imply that the ripples move. As the fluid flow perturbs the bed only to a small depth, the former segregation patterns are ``glued'' under the newer ones. The genesis of a ripple can thus be deduced from the observation of segregation patterns.

\begin{figure}[ht]
\begin{center}
\includegraphics[width=7cm]{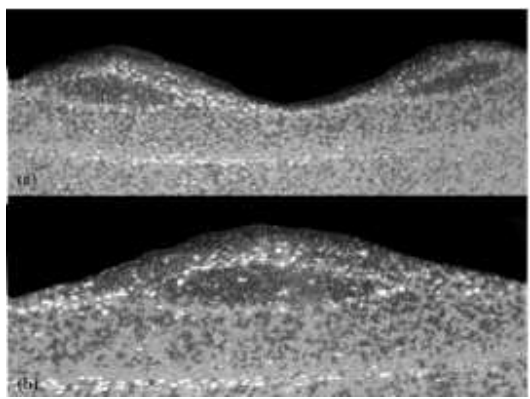}
\caption{``Cat's eyes'' pattern due to the annihilation of one small vortex ripple by two larger ones. Mixture $M1$ in Setup 1.}\label{cat}
\end{center}
\end{figure}

In most cases, the ripple coalescence does not lead to characteristic segregation patterns, by contrast to annihilation processes. Three ripples are involved in the annihilation process : a small one, situated between two larger ones. The vortex created by the larger ripples ``divides'' the smaller ripple into two parts which feed these larger ripples. The result is an ovoid-like region of larger grains surrounded by smaller grains inside each large ripple [see Figure~\ref{cat}]. This ``cat's eyes'' configuration may disappear after very long times because of new coalescence, annihilation or ripple drifts. Moreover, the segregation in volume described in the previous section may cause a ``blurring'' of the patterns because of internal convective motions.

\subsection{Sand-bed genesis}

\begin{figure}[ht]
\begin{center}
\includegraphics[width=7cm]{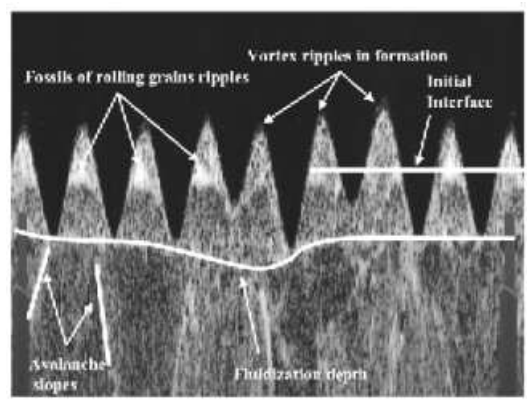}
\caption{History of the sand ripples (Mixture 2 in Setup 1) : the vertical scale is stretched ten times with respect to the horizontal. The initial preparation parameters correspond to the case of Figure 3c.} \label{hist}
\end{center}
\end{figure}

In the light of the previous sections, the {\it a-priori} observation of a sand-bed may give us much information on its history. Figure~\ref{hist} presents a sand/bed covered by vortex ripples. We can analyse this picture from bottom to top, i.e. from past to present.

First, one can see inclined lines of small (white) grains in half the bottom part of the bed. These lines correspond to the avalanche slopes of the sand and have been created when the channel has been poured. A bit higher, the avalanche lines are stopped by a nearly horizontal line corresponding to the depth of fluidization that occurred when the interface has been flattened. Higher again, the ripple bulks are mainly composed of small grains. In the case of an initially homogeneous mixture, a nearly homogeneous mixture is expected here, and then, a layer of small grains under a layer of larger grains. What we see here are fossils of rolling grain ripples built with small grains. The interface preparation was such that hydrodynamic segregation occured, moving the smaller grains to the top. Those rolling-grain ripples have been glued in the bed. Eventually, larger grains are found on the top of the ripples.
 

\section{Summary}
We have studied experimentally the formation and evolution of ripples in binary granular sand beds submitted to oscillating water flows. The importance of a well-defined protocol for the interface preparation has been demonstrated. The influence of the experimental parameters on the initial wavelength of rolling-grain ripples has been studied. In the case of binary mixtures, the wavelength seems to behave in an intermediate way between separated species. The phase segregation encountered in vortex ripples is a segregation in volume and results from the dilatancy caused by the fluid flow over the ripples. Some exotic segregation patterns have been shown and applications to geological science has been proposed.

\section*{Acknowledgments}
The experiments were financially supported by the ACI ``Jeunes chercheurs'' contract n$^\circ 2314$. HC is financially supported by the FRIA (Brussels, Belgium) and also supported through the ARC contract n$^\circ 02/07-293$. The authors are thankful to D. Vallet, O. Brouard and C. Baradel for technical help. They also thank P.-Y. Lagr\'ee for helping in the numerical calculations, as well as E. Guyon, H. Herrmann, \& N. Vandewalle \& L. Trujillo for fruitful discussions.

\end{document}